\documentstyle[aps]{revtex}

\def\be{\begin{equation}}
\def\ee{\end{equation}}
\def\ba{\begin{eqnarray}}
\def\ea{\end{eqnarray}}

\newcommand{\eq}{\begin{equation}}
\newcommand{\eqx}{\end{equation}}
\newcommand{\eqn}{\begin{eqnarray}}
\newcommand{\eqnx}{\end{eqnarray}}
\newcommand{\f}[2]{\frac{#1}{#2}}

\newcommand{\cor}[1]{\left\langle{#1}\right\rangle}

\newcommand{\lra}{\longrightarrow}
\newcommand{\DD}{{\cal D}}
\newcommand{\al}{\alpha}

\newcommand{\gm}{\gamma}
\newcommand{\dl}{\delta}
\newcommand{\xd}{\dot{X}}
\newcommand{\vb}{\bar{v}}
\newcommand{\ub}{\bar{u}}
\newcommand{\tr}{\mbox{\rm tr }}
\newcommand{\sg}{\sigma}
\newcommand{\taut}{\tilde{\tau}}
\newcommand{\qb}{\bar{q}}
\renewcommand{\th}{\theta}
\newcommand{\ph}{\phi}

\newcommand{\bra}[1]{\left\langle {#1} \right|}
\newcommand{\ket}[1]{\left| {#1} \right\rangle}

\newcommand{\vectwo}[2]{%
\left(\begin{tabular}{c}
$#1$  \\
$#2$
\end{tabular}\right)}

\newcommand{\arr}[4]{%
\left(\begin{tabular}{cc}
$#1$ & $#2$ \\
$#3$ & $#4$
\end{tabular}\right)}

\newcommand{\scr}{\scriptstyle}
\newcommand{\Tau}{{\cal T}}
\newcommand{\alef}{\al_{eff}'}
\renewcommand{\AA}{{\cal A}}

\newcommand{\rr}[4]{#1, {\it #2 \/}{\bf #3} #4}

\begin{document}

\title{Reggeon exchange from $AdS/CFT$
}

%\author{R. A. Janik\thanks{%
%M. Smoluchowski Institute of Physics, 
%Jagellonian University, 
%Reymonta 4, 30-059 Cracow, Poland;\newline
%e-mail:ufrjanik@if.uj.edu.pl} and R. Peschanski\thanks{%
%CEA, Service de Physique Theorique, CE-Saclay, F-91191 Gif-sur-Yvette Cedex,
%France;\newline  e-mail:pesch@spht.saclay.cea.fr}}

\author{R.A. Janik$^a$\thanks{e-mail:ufrjanik@if.uj.edu.pl} and
R. Peschanski$^b$\thanks{e-mail:pesch@spht.saclay.cea.fr}}

\address{%
$^a$ M. Smoluchowski Institute of Physics, 
Jagellonian University,
Reymonta 4, 30-059 Cracow, Poland\newline
$^b$ CEA, Service de Physique Theorique, CE-Saclay, F-91191
Gif-sur-Yvette Cedex, France}

\maketitle

\begin{abstract}
Using the AdS/CFT correspondence in a confining background
and the worldline formalism of gauge field theories,
we compute scattering amplitudes with an exchange of quark and
antiquark in the $t$-channel corresponding to Reggeon exchange. It
requires going beyond the eikonal approximation, which was used when studying 
Pomeron exchange. The wordline path integral is 
evaluated through the determination of minimal surfaces and their boundaries
by the saddle-point method at large gauge coupling $g^2N_c.$
We find a Regge behaviour with linear Regge trajectories. The slope is
related to the $q\bar q$ static potential and is four times the Pomeron
slope obtained in the same framework. A contribution to the  intercept, related 
to
the 
L\"uscher term, comes
from 
the fluctuations around the minimal 
surface.
\end{abstract}

\bigskip

\section {Introduction} 

The aim of our study is to continue our investigation using the AdS/CFT 
correspondence \cite{ma98,ma99} for application
to the description \cite{us1,us2,fluct} of high energy
scattering in the strong coupling (nonperturbative) regime of QCD.
As is well known, the high energy scattering amplitudes may be 
phenomenologically described by the exchange of states which
correspond to singularities in the complex angular momentum in the
crossed channel. These Regge singularities are of two types. Pomeron
exchange is leading in   elastic scattering at high energy and corresponds to 
vacuum quantum numbers while Reggeon
exchanges are subleading  and involves {\it a priori} various quantum number 
exchanges. The high energy
behaviour of amplitudes $s^{\al_{P,R}(t)}$ is characterized by 
universal Pomeron and Reggeon
trajectories, whose parameters are obtained from the analysis of
various soft hadronic reactions. Indeed they differ both in intercept
and slope, as shown by typical values \cite{la92} : $\al_P(t)\approx
1.08+0.25\  t$ and $\al_R(t)\approx .55 +.93\ t$, for the dominant
trajectories. Qualitatively
the difference in intercept between Pomeron and Reggeon trajectories
can be related in QCD to the effect of the exchange of wee partons
\cite{feyn}, which are gluons or $q\bar q$ for the Pomeron  and valence 
quarks for the Reggeons. However  a more precise theoretical 
determination from quantum field theories is made
difficult by its nonperturbative character.

Previous theoretical approaches to this strong coupling problem have
mainly focused on the Pomeron trajectory \cite{Nacht,Nachtr,KKL,SZN}. One 
technical reason seems to rely on
the applicability of the eikonal approximation in non-perturbative QCD 
calculations for
quark-(anti-)quark scattering with vacuum quantum number exchange for
which the   
quark propagators are  essentially
mapped
onto  Wilson lines following the straight line classical quark
trajectories.    

Using the same approximation in the framework of  the AdS/CFT correspondence  
\cite{us1,us2,fluct,SRZ}, the elastic $qq$ or $q\qb$ scattering amplitude
is given by  the 
expectation value of a Wilson loop, which is related to a minimal
surface problem in an appropriate confining version of the AdS/CFT
correspondence e.g. the AdS Black Hole (AdS BH) geometry
\cite{wittenbh}. The physical
amplitude is obtained through analytic continuation from Euclidean to
Minkowski signature. The classical approximation gives rise to the 
determination of the Pomeron slope and to a unit
intercept, while fluctuations around the minimal surface
have the effect of 
adding  a shift to the intercept above one. The values obtained in
Ref. \cite{us2,fluct} agree well with the Pomeron trajectory.
  
In this paper we would like to extend this approach to the
investigation of amplitudes which are
mediated by the exchange of a Reggeon. 
We therefore have to consider the exchange of quarks
in the $t$ channel for which  the eikonal approximation is obviously no longer
valid. This necessitates the introduction of new tools to express the
scattering amplitudes in terms of quantities computable using the AdS/CFT
correspondence. Our method is to use the so-called worldline formalism
\cite{Feyn,Schub} with fermionic spin factors
\cite{Polsf,Korsf1,Korsf2,NRZ} 
which expresses the 4-point function corresponding to the interaction of four 
$q\bar q$ states in terms of a path integral over
quark trajectories in spacetime. We will again use AdS/CFT to compute the 
Wilson
loop VEV along these trajectories. Then we will perform the path
integral over the trajectories using semiclassical approximation.
An important point is that  we  use the standard
confining backgrounds of AdS/CFT, and thus neglect the effect of
fermion loops.

The relation between scattering amplitudes and the 4-point function
involves the LSZ reduction and the knowledge of the wavefunction of
the asymptotic states. For sake of simplicity we will assume in the
present paper that the wavefunction is just the product of free
spinors. The factors due to the propagation of the asymptotic states
towards and from the interaction region (which are truncated by LSZ reduction)
are not included in our calculation. A more refined analysis would
have to discuss  the structure of  the confined 
$q\qb$ bound
 states which is beyond the scope of our study.

The paper is organized as follows. In section II the different
ingredients of the worldline formalism, namely quark trajectories,
mass term, spin factor and Wilson loop VEV, are introduced. In section III we
explain how to perform averaging over gauge fields for the Wilson loop VEV in 
the AdS BH
background. In section IV we introduce the geometry of the minimal surface and 
its boundaries and  compute the relevant spin factor, while in section V we 
evaluate the remaining path
integral over the boundaries and give the final results, leading to the 
determination of the Reggeon trajectory. In section VI we give a
summary of our results
and discussion. An appendix is devoted to the mathematical derivation  
of a generic spin factor for trajectories embedded in a 3-dimensional
subspace of the 4-dimensional spacetime.

\bigskip

\section{Scattering amplitudes with Reggeon exchange}

\bigskip \input epsf \vsize=8.truecm 
%\hsize=20.truecm
 \epsfxsize=8.cm{%
\centerline{\epsfbox{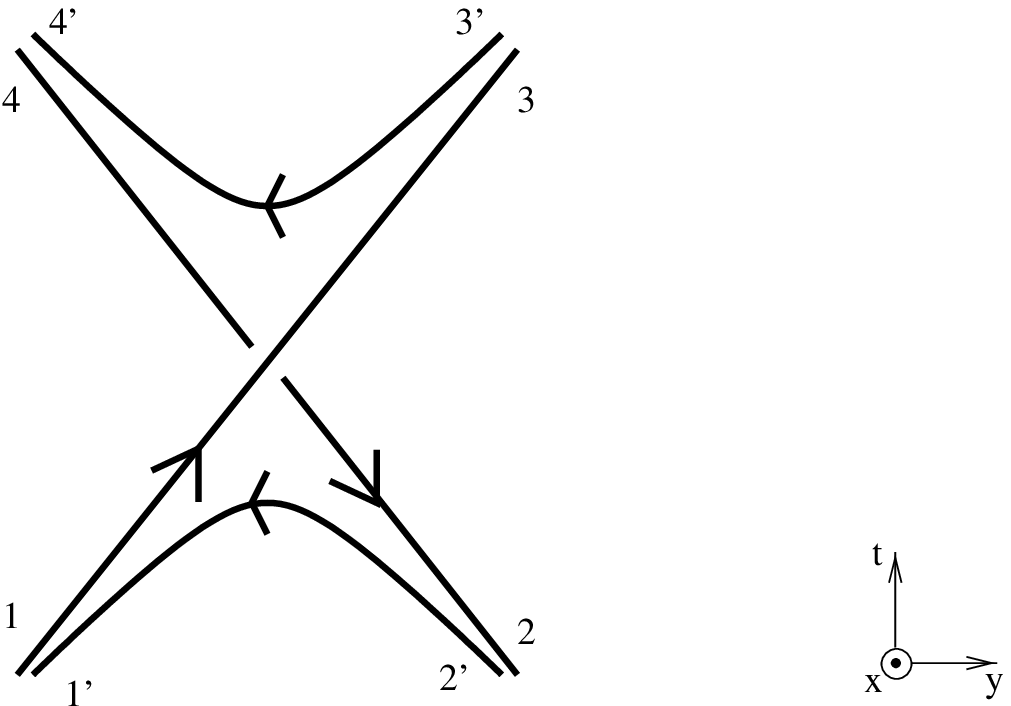}}} {\bf Figure 1}

{\it Spacetime picture of a meson-meson scattering process  mediated by
Reggeon exchange.
The impact parameter axis is perpendicular to the
longitudinal $t-y$ plane.}

\bigskip

Let us consider a meson-meson scattering process
\eq
\label{e.scat}
(11')+(22') \lra (33')+(44'),
\eqx
where the continous lines 1-3, 2-4  correspond to spectator quark and
antiquark, 
while the dashed lines 1'-2', 3'-4' correspond to annihilated and produced
quark-antiquark pairs. The labels correspond to the initial and final
spacetime position 4-vectors that we fix for our calculation.

The spacetime picture of this process is schematically illustrated
in figure 1, where the impact parameter axis $x$ is perpendicular to the
longitudinal $t-y$ plane. Note that the impact parameter is defined w.r.t. the 
spectator quark asymptotic trajectories.

The amplitude 
corresponding to the scattering process (\ref{e.scat}) can be
schematically written as 
\eq
\label{e.ampgen}
\cor{ <\!\!out|
S_F(3',4'|\AA)\, S_F(1,3|\AA)\, S_F(4,2|\AA)\, S_F(2',1'|\AA) 
|in\!\!> }_\AA
\eqx
where $<\!\!out|$ and $|in\!\!>$ are wavefunctions for the outgoing
and incoming mesons (up to modifications due to LSZ reduction
formulae). In formula (\ref{e.ampgen}),
$S_F(X,Y|\AA)$ denotes the
full quark propagator between spacetime points $X$ and $Y$ 
in a given background gauge field configuration $\AA$,
while the correlation function $\cor{\ldots}_\AA$ stands for averaging
over these configurations. 

Let us first perform the calculations for the above
scattering amplitude rotated into Euclidean space. 
%After completing the calculation 
Then we will rotate it back to Minkowski space \cite{Megg} using the
substitutions
\eqn
t &\lra& i t\\
\th &\lra& -i\chi \sim -i \log s
\eqnx
where $\th$ is the angle between the asymptotic straight lines $1\to
3$ and $2\to 4$. 
%In addition it is convenient to make a Fourier
%transform to impact parameter space and use
%It is convenient to use expressions for the
%fermionic propagators in position space. 
%The most useful expression, from the point of view of our formalism
%is to use 
In impact parameter space we use 
the worldline expression for the (Euclidean) fermion propagator in a
background gauge field $\AA = A^C_\mu(X^\mu)$ as a path
integral over classical trajectories \cite{Polsf,Korsf1,Korsf2,NRZ}:
\eq
\label{e.s}
S_F(X,Y|\AA)= \int_0^\infty dT e^{-m T} \int \DD X^\mu(\tau) 
\dl(\xd^2\!-\!1)\, I[X^\mu(\tau)]\, P e^{i\int A_\mu(X(\tau))\cdot \xd^\mu 
d\tau} 
\eqx
Here the path integral is over trajectories $X^\mu(\tau)$ joining $X$
and $Y$, parametrized by $\tau\in (0,T)$. Because of the delta
function, $T$ is also the total length of the trajectory. 
The quark mass dependence appears in the first exponential.
The colour and gauge field dependence is encoded in the
(open) Wilson line along the trajectory $P e^{i\int A_\mu(X(\tau))\,
\xd^\mu d\tau}$, while the spin 1/2 character 
of the quark is responsible for the appearance of the spin factor:
\eqn
\label{e.sfdef}
I[X^\mu(\tau)]&=&P \prod \f{1+\xd^\mu \gm_\mu}{2}
= \lim_{N\to \infty} 
\f{1+\xd^\mu(T) \gm_\mu}{2} \cdot \ldots \cdot \f{1+\xd^\mu(\f{2}{N}T)
\gm_\mu}{2} \f{1+\xd^\mu(\f{1}{N}T) \gm_\mu}{2}
\eqnx
where the second equality gives a suitably regularized definition of the 
infinite 
product
along the trajectory $X^\mu(\tau)$.
Note that each of the $N$ factors in this expression is
a projector due to the fact that $\xd^2=1$.
This spin factor was first formulated for D=3 in \cite{Polsf} and later
for arbitrary D in \cite{Korsf1,Korsf2}. 
In practice it was computed explicitly in D=2 and D=3, but not in
general for $D>3$. One of our goals is to compute it for the
configuration of figure 1 in $D=4$ spacetime. The calculation is
presented in the Appendix. 

Let us first make two comments about the calculation of (\ref{e.ampgen}).
   
i) Since the initial and final mesons are colour singlets, the four
Wilson lines   close to
form a single Wilson loop, and the gauge
averaging factorizes out of the expression:
\eq
\cor{\tr P e^{i\int_C A}}_\AA
\eqx
where the contour $C$ follows the quark trajectories  $1\to 3' \to
4' \to 2' \to 1'$ (following the contours sketched  on figure~1).

ii) The spin factor matrices
\eq
I[1\!\to\! 3]_{\al_1\al_3}\, I[4\!\to\! 2]_{\al_4\al_2}\,
I[2'\!\to\! 1']_{\al_{2'}\al_{1'}}\, I[3'\!\to\! 4']_{\al_{3'}\al_{4'}} 
\eqx
are contracted with the  initial and final spinor
wavefunctions like $u_{\al_1}(p_1)\vb_{\al_{1'}}(p_1),$ corresponding to a 
simple approximation for the wave-functions of the external mesons as
mentioned in the introduction.

\bigskip

\section {Averaging over the gauge fields --- The AdS Black Hole
background} 

According to the AdS/CFT correspondence the expectation value of a
Wilson loop is given by the string partition function with the
condition that the AdS string ends on the curve $C$ which is
placed on the boundary of the AdS geometry, which represents the
physical spacetime \cite{loopsads}. Using the saddle point
approximation (at large $g^2N$ gauge coupling) 
gives the formula:
\eq
\label{e.wilson}
\cor{W(C)}\approx Fluctuations(C) \cdot e^{-\f{1}{2\pi \al'} Area(C)}
\eqx
where $Area(C)$ is the minimal area of a string worldsheet evaluated in the
{\em curved} geometry of the background. This background has to be
choosen dual to the appropriate gauge field
theory. $\al'$ is the AdS string tension.
%This exponential factor corresponds to inserting the saddle
%point configuration (minimal surface) into the string action, while 
The prefactor $Fluctuations(C)$ is the contribution of quadratic
fluctuations around the minimal surface. 

In this paper we will use the confining Black Hole (BH)
background proposed in \cite{wittenbh}.
Let us recall the main features of the evaluation of the expectation
values of Wilson loops in this geometry.

When the Wilson loop is large enough (in comparision to the horizon
radius, which sets the confinement scale of the theory), the minimal
surface is concentrated near the horizon and
we can use the flat space approximation.
Then (\ref{e.wilson}) takes the form
\eq
\label{e.wilsonfl}
\cor{W(C)}=Fluctuations(C) \cdot e^{-\f{1}{2\pi \alef} Area_{FLAT}(C)}
\eqx
where now the minimal surface area is evaluated in the flat metric and
the parameter $\alef$ involves
%has been substituted to the AdS string tension and is 
%characterized by  
a scale related to the BH metric near
the horizon. The version of AdS background that we are using does not
allow to determine $\alef$ from first principles.
It can be directly related, however, to the static quark-antiquark
potential obtained from a rectangular Wilson loop of size $T \times L$: 
\eq
\label{e.wtl}
\cor{W(T\times L)}\equiv e^{TV(L)}=
e^{T\left[ -\f{L}{2\pi \alef} + n_\perp \cdot \f{\pi}{24} \f{1}{L} + \cdot 
\cdot \cdot\right]}\ ,
\eqx
where $T,L \ and \ T/L\gg 1.$

The first term in the potential comes from the area of the rectangle,
while the second one is due to fluctuations
\cite{luscher,alvarez,frts,arvis,kinar}.
Note that, when calculating the fluctuations around 
a minimal surface near the horizon in the
BH backgrounds there could be  $n_\perp=7,8$ massless bosonic modes 
\cite{kinar}. 
In general we will keep  $n_\perp$ as a free parameter in order to
accomodate other geometric realizations of a confining gauge theory. 
It is at this stage that the string picture from AdS/CFT may deviate
from the old `effective string' picture of QCD. The $n_\perp$ need not
be necessarily equal to $4-2=2$. The ambient number of
effective dimensions in the dual string theory may be indeed higher.
The 
question of its interpretation in terms of gauge field theory collective 
degrees of freedom 
remains an interesting open question \cite{schur}.

Within our approximation we will just use the two parameters $\alef$ and
$n_\perp$ which parametrize the behaviour of string theory in the
confining geometry. It is interesting to note that even in the absence
of an {\em ab initio} theoretical determination of these parameters,
they can be obtained from e.g. lattice QCD calculations and used to
predict the behaviour of scattering amplitudes at high energy.

\section{Evaluation of the scattering amplitudes}

We will now proceed to evaluate the scattering amplitude. In this
section we will just concentrate on the classical configurations
leaving the evaluation of the contribution of quadratic fluctuations to the
following section.

In the previously known nonperturbative calculations of scattering
amplitudes using Wilson line/loop formalism, it was crucial to
consider the eikonal approximation for the quark trajectories
\cite{Nacht,Nachtr,SZN,us1}. This is
very well justified for elastic amplitudes (Pomeron trajectories),
where all quarks are effectively spectators i.e. their spacetime
trajectories are supposed not to be deflected by the gluon field.
On the contrary, the Reggeon exchange amplitude implies that two quark
lines (i.e. $2' \to 1'$ and $3' \to 4'$ in figure 1) are exchanged in 
the $t$ channel and thus the corresponding
propagators cannot be described within the eikonal approximation. 

\subsection*{Spectator quarks}

The eikonal approximation is expected still to be valid for the
spectator quarks (i.e. $1 \to 3$ and $2 \to 4$ in figure 1).  
It will be convenient to assume that the spectator quarks  are heavy, 
while the exchanged quarks are light.
Hence the spectators just 
follow straight line trajectories, while the path integrals of
noneikonal type have to be done w.r.t the trajectories $2' \to 1'$ and
$3' \to 4'$. 

The spin factors for the spectator quarks are
simple. Indeed, all the terms in the product (\ref{e.sfdef}) are
identical as $\xd$ is constant along the trajectory. 
Since they are projectors, the product reduces to a
single term. Therefore the spin factor is just
\eq
\label{e.spproj}
\f{1+\xd^\mu \gm_\mu}{2} \equiv \f{1+\f{\hat{p}_i}{|p_i|}}{2} \quad i=1,2
\ .
\eqx
When we act with this projector on the spinor $u(p_i)$, we have
$\hat{p}_iu(p_i)=|p_i| u(p_i)\equiv m u(p_i)$, so that e.g.
$I[1\!\to\!3]u(p_1)=u(p_1)$. Contracting this with the outgoing spinor
will give the prefactor (in the forward direction)
\eq
\ub_{\al_3}(p_1)u_{\al_1}(p_1)=\dl_{\al_3 \al_1}
\ .
\eqx

\subsection*{Exchanged quarks}

Let us consider the path integral expression for (\ref{e.ampgen})
after inserting the four quark propagators (\ref{e.s}). Integrating
first over the gauge field configurations (for fixed trajectories),
the amplitude can be be schematically rewritten as
\eqn
\label{e.amp}
\int {\cal D} X^\mu(\tau) \biggl\{ \dl(\xd^2-1) \cdot 
(\mbox{Spin factors})\cdot \cor{W(1\to 3' \to 4' \to 2' \to 1')}_\AA
\ 
e^{-m\cdot (Length[2' \to 1']+Length[3' \to 4'])} \biggr\} \ ,
\eqnx
where the contributions of the lengths of the spectator trajectories
are included in the implicit normalization.  
Since we assumed that the spectator quarks are heavy and  follow
straight line trajectories, the Wilson loop is formed out of two
straight lines and the trajectories $2' \to 1'$ and $3' \to 4'$.

The dominant contribution in (\ref{e.amp}) will
come from evaluating the Wilson loop expectation value 
(saddle point in $\alef$ c.f. (\ref{e.wilson})) through  the minimal
area in the AdS BH bulk having this loop as the boundary.

Let us now assume that the exchanged quarks are nearly massless. Our
main assumption is that their trajectories will be constrained to
remain on the minimal surface spanned between the two (infinite)
spectator trajectories. This is a kind of {\it softness} assumption which is 
sensible  in the classical approximation that we consider.  

The relevant minimal surface is the helicoid \cite{us2}
parametrized by
\eqn
\label{e.helt}
t &=& \tau \cos p\sg \\
y &=& \tau \sin p\sg \\
\label{e.helx}
x &=& \sg
\eqnx
where 
\eq
\label{e.defofp}
p=\th/L, \quad\quad\quad \quad \sg=-L/2 \ldots L/2 \ .
\eqx
and $L$ is the impact parameter distance.
The nearly massless exchanged quarks follow trajectories
{\em on} the helicoid, which are defined by specifying the
functions $\tau(\sg)$. These trajectories  form the upper and
lower edges of the Wilson loop and will be obtained through a
subsequent minimization. 

Averaging over the gauge fields involves, according to
(\ref{e.wilsonfl}), the calculation of the area of the piece of
helicoid between the boundaries $\pm \tau(\sg)$. 
Using the results \cite{us2} we obtain
\eqn
\label{e.area}
Area[\tau(\sg)] =\int_{-L/2}^{L/2} d\sg
\int_{-\tau(\sg)}^{\tau(\sg)} d\taut \sqrt{1+p^2\taut^2} = \int_{-L/2}^{L/2} 
d\sg 
\, \left\{\tau(\sg) \sqrt{1+p^2\tau^2(\sg)}+ \f{1}{p}\log\left( p\tau(\sg) +
\sqrt{1+p^2\tau^2(\sg)} \right) \right\}
\eqnx
while the length of the edges is given by
\eq
\label{e.length}
Length[\tau(\sg)]= \int_{-L/2}^{L/2} d\sg \sqrt{1+p^2 \tau^2(\sg)
+\left(\f{d\tau(\sg)}{d\sg}\right)^2}\ . 
\eqx

\subsection*{Spin factor}

It remains to evaluate the remaining piece of the amplitude
(\ref{e.amp}), the spin factor associated to the trajectories $2' \to 1'$
and $3' \to 4'$. Since the quark trajectories are contained in the
helicoid, they lie in a 3 dimensional hypersurface of the 4D physical 
spacetime. For these types of trajectories the spin factor simplifies
substantially. The result of the calculation presented in the Appendix
boils down to the following expression
\eq
I[\xd]=\f{1+\xd^\mu(T) \gm^\mu}{2} \f{1+\xd^\mu(0) \gm^\mu}{2} \cdot
\left(\f{1+\xd(T)\cdot\xd(0)}{2}\right)^{-1} \,
\eqx 
which depends only on the initial and final direction vectors
$\xd^\mu(0)$, $\xd^\mu(T)$, and not on the specific form of the
whole trajectory, provided it is smooth. 

If we contract the projectors with the appropriate spinors
$u(p_i),\vb(p_j)$ the projectors act like the identity after using
the free Dirac equation (recall the discussion after formula
(\ref{e.spproj})). The scalar products $\xd(T)\cdot\xd(0)$ are equal 
to $-\cos\th \to -s$, therefore in the Minkowskian high energy limit
we obtain after contraction
the factor 
\eq
\label{e.finalsf}
{\cal I}=\vb_{\al_{1'}}(p_{1})u_{\al_{2'}}(p_{2}) \cdot \ub_{\al_{4'}}(p_{2})
v_{\al_{3'}}(p_{1}) \cdot \f{1}{s^2} = -\f{1}{s} \cdot \sg^3_{\al_{1'}\al_{2'}}
\sg^3_{\al_{4'}\al_{3'}} \ . 
\eqx
%
%\eqn
%\vb_{\al_{1'}}(p_{1})u_{\al_{2'}}(p_{2}) &=& \f{p_1^\Vert}{m}
%\cdot \sg^3_{\al_{1'}\al_{2'}  } \sim
%\sqrt{s} \cdot \sg^3_{\al_{1'}\al_{2'} }  \\
%\ub_{\al_{4'}}(p_{2}) v_{\al_{3'}}(p_{1}) &=& -\f{p_1^\Vert}{m} \cdot
%\sg^3_{\al_{4'}\al_{3'} } \sim -\sqrt{s} \cdot  \sg^3_{\al_{4'}\al_{3'} } 
%\eqnx
%Finally we get from the kinematical spin factors
%\eq
%-\f{1}{s} \cdot \sg^3_{\al_{1'}\al_{2'}} \sg^3_{\al_{4'}\al_{3'}}  
%\eqx
%This spin factor calculation coincides with the standard perturbative
%expectations that the exchange of two spin $1/2$ particles will give a
%$1/s$ suppression of the amplitude. 

\section{Reggeon exchange amplitudes}

The final result is obtained through the computation of the path
integral over the trajectories of the exchanged quarks parametrized by
the function $\tau(\sg)$. Using the results of the previous section,
formula (\ref{e.amp}) can be rewritten as 
\eq
\label{e.pathint}
{\cal I}\cdot \int \DD\tau(\sg)\, Fluctuations[\tau(\sg)]\, e^{-\f{1}{2\pi
\alef} Area[\tau(\sg)] -2m 
Length[\tau(\sg)]}      \ .
\eqx
The $Area[\tau(\sg)]$ is given by formula (\ref{e.area}),
$Fluctuations[\tau(\sg)]$ is the contribution of quadratic
fluctuations of the string around the given minimal surface defined by
the trajectory $\tau(\sg),$ see (\ref{e.wilsonfl}).
$Length[\tau(\sg)]$ appears in formula (\ref{e.length}). Note that the
trajectory-independent spin factor ${\cal I}$ obtained in (\ref{e.finalsf}) has 
been
factored out. 

Let us perform this integral by saddle point in the limit of
$\alef$ small. As a consequence we may consider both the $m$
dependence and the fluctuation contribution as prefactors not entering
the saddle point equations. This should be true for small enough
masses $m$. We will discuss this point later on.

The Euler-Lagrange equations are
\eq
\label{e.el}
\f{\partial Area[\tau(\sg)]}{\partial \tau} =\sqrt{1+p^2 \tau(\sg)^2}=0
\ .
\eqx 
The solution is a {\em constant complex} $\tau(\sg)=\pm i/p$. Here the complex
value has to be understood in the sense of applying the steepest descent
method to the path integral (\ref{e.pathint}), and deforming the
integration contours into the complex plane.

Physically this indicates an instability of the minimal surface
problem associated with the scattering process. This is reminiscent of
a similar phenomenon encountered in the semiclassical study of high energy 
scattering
of asymptotic open string states in flat spacetime c.f. Ref. \cite{gross}.

Substitution of the classical solution $p\tau(\sg)=-i$ into
(\ref{e.pathint}) gives a non vanishing contribution from the
logarithm:
\eq
\label{e.spa}
e^{-\f{1}{2\pi\alef} Area[-i/p]} = e^{-\f{i L^2}{4\alef \th}} \lra 
e^{-\f{ L^2}{4\alef \chi}} 
\eqx
after analytical continuation to
Minkowski space. 

\subsection*{Fluctuations}

Let us evaluate the contribution of quadratic fluctuations of the
string worldsheet around the minimal surface defined by the saddle point
trajectories (\ref{e.el}). 

Since the classical trajectory is constant (albeit complex), we will
calculate the result for fixed $\tau(\sg)=\Tau$ for real $\Tau$, and then
analytically continue to $\Tau=-i/p$. 

The fluctuation determinant for the case of a helicoid bounded by two
helices with $\tau=\pm \Tau$ has already been calculated \cite{fluct}. Let us
briefly recall the basic steps. First one replaces
the variable $\tau$ in (\ref{e.helt})-(\ref{e.helx}) by
\eq
\rho=\f{1}{p}\log(p\tau + \sqrt{1+p^2 \tau^2}) \ .
\eqx
In the variables $\rho$, $\sg$ the induced metric on the helicoid is
conformally flat i.e. 
\eq
g_{ab}=(\cosh^2 p\rho)\ \dl_{ab} \ .
\eqx
Therefore,
since string theory in the AdS background is {\em critical}, we may
perform the 
calculation for the conformally equivalent  flat metric $g_{ab}=\dl_{ab}$. This 
reduces to a
calculation of the fluctuation determinant for a rectangle of size
$a\times b$ where
\eqn
a &=& L\\
b &=& \f{2L}{\th} \log \left( p\Tau+\sqrt{1+p^2 \Tau^2} \right)
\eqnx 
We assume furthermore that the quadratic bosonic fluctuations are
governed by the Polyakov action, as is indeed the case for string
theories on AdS backgrounds. 
The result is just equivalent to the L\"uscher term
computation
(c.f. (\ref{e.wtl})), and for high energies (after continuation to
Minkowski space $a/b ={\cal O}(\log s) \gg 1$) we obtain  
\eq
\label{e.fluct}
Fluctuations(\tau(\sg)\equiv\Tau)=
%\exp \left( n_\perp \cdot
%\f{\pi}{24} \cdot \f{a}{b} \right)=
\exp \left( n_\perp \cdot \f{\pi}{24} \cdot
\f{\th}{2 \log\left( p\Tau +\sqrt{1+p^2\Tau^2} \right)} \right)
\ .
\eqx
Analytically continuing this expression to the saddle point $\Tau=-i/p$ gives
\eq
\label{e.fluctsp}
Fluctuations(\tau(\sg)=-i/p)=e^{\f{n_\perp}{24}\log s}=s^{\f{n_\perp}{24}}
\eqx

Putting together the different components contributing to
(\ref{e.pathint}), namely the spin factor (\ref{e.finalsf}), the
classical minimal area contribution (\ref{e.spa}) and the fluctuation
determinant (\ref{e.fluctsp}), we get for the amplitude in impact parameter 
representation
\eq
\label{e.ampimp}
s^{-1+\f{n_\perp}{24}} \cdot  e^{-\f{L^2}{4\alef \log s}} \cdot
\dl_{\al_1\al_3}\dl_{\al_2 \al_4} \sg^3_{\al_{1'}\al_{2'}}
\ .
\sg^3_{\al_{4'}\al_{3'}}\ .   
\eqx
A Fourier transform to momentum space gives the following result for the 
scattering amplitude:
\eq
\label{e.traj}
s^{\f{n_\perp}{24}+\alef t} \cdot
\dl_{\al_1\al_3}\dl_{\al_2 \al_4} \sg^3_{\al_{1'}\al_{2'}}
\sg^3_{\al_{4'}\al_{3'}}   
\eqx
i.e. a linear Regge trajectory
\eq
\label{e.lintraj}
\al(t)=\f{n_\perp}{24}+\alef t
\ .
\eqx
Note that we neglected in (\ref{e.traj}) possible logarithmic
prefactors which are not under control at this stage of our approach.

\subsection*{Relation with Pomeron exchange}

Let us compare our resulting amplitude with the one obtained for
Pomeron exchange for which we obtained the trajectory \cite{us2,fluct}
\eq
\al_P(t)=1+\f{n_\perp}{96}+\f{\alef}{4} t
\ .
\eqx
The first observation is that in both cases, the slope is determined
by minimal surface solutions through the logarithmic contribution in
the helicoid area. The factor 4 difference in the slope
comes from the specific saddle point path integral over the exchanged quark
trajectories (for Reggeon exchange). It is interesting to note that
this theoretical feature is in agreement with the phenomenology of
soft scattering. Indeed once we fix the $\alef$ from the
phenomenological value of the static $q\qb$ potential ($\alef\sim 0.9\,
GeV^{-2}$) 
we get for the slopes $\al_R=\alef\sim 0.9\, GeV^{-2}$ and
$\al_P=\alef/4\sim 0.23\, GeV^{-2}$ in good agreement with the observed
slopes, c.f. section I.

The second feature is the relation between the Pomeron and Reggeon
intercepts. At the classical level of our approach these are
respectively 1 and 0. Note that this classical piece is in
agreement with what is obtained  from simple exchanges of two
gluons and quark-antiquark pair, respectively, in the $t$ channel. The 
fluctuation (quantum) contributions to the Reggeon and Pomeron are
related by a factor of 4.   
Adding both classical and fluctuation contributions gives an estimate
which is in qualitative agreement with the observed intercepts.  
For $n_\perp=7,8$ one gets $1.073-1.083$ for the Pomeron and $0.3-0.33$
for the Reggeon. 
This result is below the intercepts of around $0.5$ observed for the
dominant Reggeon trajectories. We will discuss some possible sources
of this discrepancy in the summary.

An interesting feature of the formulae for the area (\ref{e.area}) of
the helicoid and fluctuations (\ref{e.fluct}) around it, is the
key role of the logarithmic term. 
This gives rise to the possibility of 
additional contributions from passing onto a different Riemann sheet
($log \to log+2\pi i k$)
in the course of performing analytical continuation from Euclidean to
Minkowski space \cite{us2,fluct}. 
The amplitude in impact parameter space (\ref{e.ampimp}) will then pick
up new multiplicative factors:
\eq
\label{e.rp}
e^{-\f{L^2}{4\alef \log s}} \cdot e^{-k\f{L^2}{\alef \log s}} \ .
\eqx
This can be interpreted (for $k>0$) as $k$-Pomeron exchange 
corrections to a single Reggeon exchange. Indeed the slope of the
trajectory obtained from Fourier transform of formula (\ref{e.rp}) is
just the one expected from such contributions. 
These are well known to
describe absorptive corrections in Regge phenomenology. 
Note however that the intercept decreases with $k$ like
$\f{n_\perp}{24(1+4k)}$ instead of growing as expected from usual absorptive
correction models.

\section{Summary and Discussion}

In the present paper we proposed a method to study Reggeon exchange
amplitudes starting from basics of nonperturbative gauge theory. 
Since the process involves the exchange of quarks between the incoming
states the eikonal approximation, which is useful for Pomeron
exchange, is no more valid and new tools have to be used.
We consider then the worldline formalism which relates the scattering
amplitude to a path integral over quark trajectories involving spin
factors and a Wilson loop, both depending on the spacetime quark trajectories, 
which enter in the definition of the functional integral.    

We give a semiclassical evaluation of the Wilson loop VEV using the 
AdS/CFT correspondence for a confining theory (AdS Black
Hole). The appropriate minimal surface in the bulk turns out to be a
portion of a helicoid whose boundaries, determined by minimization,
are the exchanged quark trajectories, analytically continued in the complex 
plane.
We prove that for such case, where the trajectories are contained in a
3-dimensional subspace of 4-dimensional spacetime, the spin factor
depends only on the initial and final momenta and thus factor out of the 
functional integral. In this case this leads to
a $1/s$ factor corresponding the Regge intercept 0, which is what is
expected from two spin $1/2$ exchanges.

In this framework we find a Regge behaviour with linear trajectory
with slope $\alef$, directly related to the static $q\qb$
potential. Interestingly enough from a phenomenological point of view, this is 
exactly four times the slope of the Pomeron trajectory obtained in
\cite{us2,fluct} within the same approach.

The contribution of worldsheet fluctuations around the minimal surface
for fixed boundaries gives rise, after analytical continuation to
Minkowski space, to an intercept increment equal to $n_\perp/24$ where
$n_\perp\sim 7,8$ is the number of transverse massless fluctuation
modes of the string in the 10 dimensional bulk. This value
is directly related to the L\"uscher term coefficient in the static
potential. It is four times the increment of the Pomeron intercept
above 1 found in the same approach \cite{fluct}.

Let us add a few comments. The minimization procedure leading to
(\ref{e.el}) can be extended by including the mass term in the saddle
point determination also leading to a constant trajectory solution
$\tau(\sg)=constant$. We checked that this constant goes to the value
obtained in section 5 when $m \to 0$. However if $m$ is not small
enough we do  not expect that the minimal surface would  remain
helicoidal, since it could be deformed by mass terms involved in the 
minimization.

In addition, apart from the fluctuations around the minimal surface with
fixed (saddle point) boundaries, one expects also a contribution from
fluctuation of the boundary trajectories, both inside the helicoid
surface, as well as outside. This is a more involved problem which
deserves to be studied in the future.

We did not study in detail the helicity structures of the
amplitudes. It is important since in the real world they play a r\^ole in 
distinguishing different Reggeon trajectories like $\rho$ and $\pi$
exchanges differing in the intercepts. In any case
we expect no difference in the Regge slopes which are obtained
independently of the spin factors. Our result, using only simplifying 
assumptions 
 on the asymptotic wave functions could correspond to a combination of 
different 
Regge trajectories. 

Finally, an interesting feature of AdS/CFT is the
possibility of exploring the transition to shorter distances, where
the curved background geometry will start to play an important role
modifying the minimal surfaces and thus the behaviour of scattering
amplitudes.
 
{\it Note added:} While completing this paper, a  study of high energy 
scattering 
in the AdS/CFT framework \cite {polch}, investigating  different processes and 
using  different tools, has appeared.  

\subsection*{Acknowledgements}
We thank A. Bialas and G. Korchemsky for interesting discussions and useful 
remarks.
RJ would like to thank the SPhT Saclay for hospitality during the 
course 
of this work. RJ was supported in part by KBN grant 2P03B01917.

\pagebreak

\subsection*{Appendix --- Evaluation of the spin factor}

In this appendix we will evaluate the spin factor in 4D for a
trajectory which is contained in the 3D hypersurface $(t,x,y)$.
We will use the standard basis of gamma matrices, rotated by $i$ to
Euclidean space:
\eq
\gm^0=\arr{1}{0}{0}{-1} \quad\quad\quad\quad
\gm^k=\arr{0}{i\sg^k}{-i\sg^k}{0}
\ .
\eqx
Following \cite{Korsf1}, we  evaluate the product of projectors
\eq
I[\xd]=\prod_{i=0}^N \f{1+\xd^\mu(\tau_i) \gm^\mu}{2} \quad ; \quad\quad\quad
\tau_i=i\f{T}{N} \ .
\eqx
Parametrizing the velocity by
\eq
\xd^\mu=(\dot{t},\dot{x},\dot{y},\dot{z})=
(\sin\th\cos\ph, \sin\th\sin\ph, \cos\th,0) \ ,
\eqx
the projector has the following form
\eq
\f{1+\xd^\mu \gm^\mu}{2} = \f{1}{2}\scriptstyle{
\left(\begin{array}{cccc}
\scr 1+\sin\th\cos\ph & 0 & 0 & \scr \cos\th+i\sin\th\sin\ph \\
0 & \scr 1+\sin\th\cos\ph & \scr -\cos\th+i\sin\th\sin\ph & 0 \\
0 & \scr -\cos\th-i\sin\th\sin\ph & \scr 1-\sin\th\cos\ph & 0 \\
\scr \cos\th-i\sin\th\sin\ph & 0 & 0 & \scr 1-\sin\th\cos\ph
\end{array}\right)}
\ .
\eqx
We see that this matrix is a direct sum of two $2\times 2$
matrices which are by themselves projectors. We may write it as
\eq
\label{e.decomp}
\f{1+\xd^\mu \gm^\mu}{2} =
\ket{\th,\ph,1}\ket{\th,\ph,2}\bra{\th,\ph,1}\bra{\th,\ph,2}\ ,
\eqx
where the 2-dimensional vector $\ket{\th,\ph,1}$ (respectively
$\ket{\th,\ph,2}$) is embedded in the $1\!-\!4$ (resp. $2\!-\!3$)
2-dimensional subspace of
the 4-dimensional spinorial representation. Explicitly, these vectors
are given by:
\eq
\ket{\th,\ph,1}= \f{1}{\sqrt{2(1-\cos\ph \sin\th)}}   
\vectwo{\cos\th+i\sin\th\sin\ph}{1-\sin\th \cos\ph}
\eqx 
and
\eq
\ket{\th,\ph,2}= \f{1}{\sqrt{2(1-\cos\ph \sin\th)}}   
\vectwo{-\cos\th+i\sin\th\sin\ph}{1-\sin\th \cos\ph}
\eqx 
Thanks to the decomposition (\ref{e.decomp}) the spin factor can now
be easily rewritten as
\eq
\label{e.spinf}
\ket{\th_N,\ph_N,1}\ket{\th_N,\ph_N,2}\bra{\th_0,\ph_0,1}\bra{\th_0,\ph_0,2}
\cdot \Omega
\eqx
where the {\em scalar} $\Omega$ is given by
\eq
\Omega=\prod_{i=0}^{N-1} \cor{\th_{i+1},\ph_{i+1},1|\th_{i},\ph_{i},1}
 \cor{\th_{i+1},\ph_{i+1},2|\th_{i},\ph_{i},2}
\ .
\eqx
For {\em smooth} trajectories $\Omega$ is equal to 1. Indeed we have
\eqn
\!\!\!\!\!\! \cor{\th+d\th,\phi+d\phi,1|\th,\ph,1}&=&  1-i
\f{\sin \phi\, d\th  +\cos\phi \cos\th \sin \th\, d\phi}{2-2\cos\phi\sin\th} 
\\
\!\!\!\!\!\! \cor{\th+d\th,\phi+d\phi,2|\th,\ph,2}&=&1+i
\f{\sin \phi\, d\th  +\cos\phi \cos\th \sin \th\, d\phi}{2-2\cos\phi\sin\th}
\eqnx
where we negleted higher order terms in $d\th,d\phi$.
We see therefore that the spinors get rotated in  opposite
directions in the two 2-dimensional subspaces of the 4-dimensional
spinorial representation. Putting the two contributions together we obtain:
\eq
\cor{\th+d\th,\phi+d\phi,1|\th,\ph,1}\cdot
\cor{\th+d\th,\phi+d\phi,2|\th,\ph,2} \sim 1+{\cal O}(d\th^2,d\phi^2,
d\th d\phi) \ . 
\eqx
Thus $\Omega$ can acquire nontrivial contributions only from cusps. 

Finally let us rewrite (\ref{e.spinf}) as
\eq
I[\xd]=\f{1+\xd^\mu(T) \gm^\mu}{2} \f{1+\xd^\mu(0) \gm^\mu}{2} \cdot
\left(\f{1+\xd(T)\xd(0)}{2}\right)^{-1} \cdot \Omega_{cusps}
\eqx
in which  we used the relation
\eq
\label{e.scalprod}
\cor{\xd(T),1|\xd(0),1}\cor{\xd(T),2|\xd(0),2}=\f{1+\xd(T)\xd(0)}{2}
\eqx
where $\Omega_{cusps}$ gets contributions only
from cusps. For the case at hand, the trajectory found from solving
the Euler-Lagrange equations (\ref{e.el}) is smooth apart from two
cusps where the curves of $\tau=const$ on the helicoid meet the
straight line trajectories of spectator quarks. Since they meet at
right angles, the contribution of each cusp is just $1/2$ as can be
derived from (\ref{e.scalprod}).

\end{document}